\documentclass[a4paper,10pt]{iopart}

\usepackage[utf8]{inputenc} 
\usepackage{graphicx}  

\expandafter\let\csname equation*\endcsname\relax
\expandafter\let\csname endequation*\endcsname\relax
\usepackage{amsmath}

\usepackage{dsfont}
\usepackage{textcomp}
\usepackage[normalem]{ulem} 
\usepackage{float}
\usepackage{units}
\usepackage[format=hang, font={footnotesize, sf}, labelfont={bf}, margin=1cm, aboveskip=5pt, position=bottom]{caption}
\usepackage{placeins}
\usepackage{cite}

\newcommand{\ud}{\,\mathrm{d}}
\newcommand{\vek}[1]{\vec{#1}}

\newcommand{\nabpara}{\ensuremath{\nabla_{\parallel}}} 
\newcommand{\laplace}{\ensuremath{\nabla^2}}

\newcommand{\neut}[1]{#1}
\newcommand{\curv}[1]{#1}
\newcommand{\diff}[1]{#1}
\newcommand{\conv}[1]{#1}
\newcommand{\esta}[1]{#1}
\newcommand{\tpho}[1]{#1}
\newcommand{\fric}[1]{#1}
\newcommand{\pres}[1]{#1}

\graphicspath{{images/}{}}

\providecommand{\bibAnnoteFile}[1]{}
\providecommand{\bibAnnote}[2]{}

\newcommand{\TITLE}{Influence of plasma background on 3D scrape-off layer filaments}
\newcommand{\TITLEL}{\TITLE} 

\providecommand{\AUTHORL}{David Schworer}

\usepackage[pdftex,%
	    pdftitle=\TITLEL,%
	    pdfauthor=\AUTHORL,%
	    colorlinks,%
	    citecolor=black,%
	    filecolor=black,%
	    urlcolor=black,%
	    linkcolor=black]{hyperref}
\begin{document}

\title{\TITLE}

\author{  D.~Schw{\"o}rer${}^{a,b}$, N.~R.~Walkden${}^{b}$, H.~Leggate${}^{a}$,
  B.~D.~Dudson${}^{c}$, F.~Militello${}^{b}$, T.~Downes${}^{a}$,
  M.~M.~Turner${}^{a}$}

\address{
      ${}^{a}$Dublin City University, Dublin 9, Ireland\\
      ${}^{b}$CCFE, Culham Science Centre, Abingdon, 
      Oxfordshire, OX14 3DB, UK\\
      ${}^{c}$ York University, York, Yorkshire, YO10 5DD, UK}
\ead{david.schworer2@mail.dcu.ie}
\vspace{10pt}
\begin{indented}
\item[]June 2018
\end{indented}



\begin{abstract}
This paper presents the effect of self-consistent plasma backgrounds
including plasma-neutral interactions, on the dynamics of filament propagation.
The principle focus is on the influence of the neutrals on the
filament through both direct interactions and through their influence
on the plasma background.
Both direct and indirect interactions influence the motion of filaments.
A monotonic increase of filament peak velocity with upstream electron
temperature is
observed, while a decrease with increasing electron density is observed.
If ordered by the target temperature, the density dependence
disappears and the filament velocity is only a function of the target
temperature.
Smaller filaments keep a density dependence, as a result of
the density dependence of the plasma viscosity.
The critical size $\delta^*$, where filaments are fastest, is shifted
to larger sizes for higher densities, due to the plasma viscosity. If
the density dependence of the plasma viscosity is removed, $\delta^*$
has no temperature dependence, but rather a density dependence.
\end{abstract}

\section{Introduction}
Filaments are field-aligned non-linear pressure perturbations that
have been observed in most magnetized plasmas~\cite{dippolito11a}.
These intermittent, localized objects have a much smaller
cross-section perpendicular to the magnetic field than parallel.
In tokamaks they can carry a significant
amount of heat and particles to the first wall materials, which may cause
sputtering, thereby diluting the plasma and degrading the wall.
The plasma wall interaction can cause dust
production as well as increase tritium retention, both concerns for
ITER~\cite{roth09a}.
Further filaments contribute to the cross field transport in the
scrape-off layer (SOL), which influences the width of the SOL and
affects the power handling at the
diveror~\cite{boedo01a,militello16a,?}.
Understanding filaments with a view to predicting and controlling
them in future devices is therefore of interest.

Computer simulation of filaments were done initially in two
dimensions~\cite{yu03a,yu06a,garcia06a}.
In 2D simulations closures are needed due to the lack of resolution in
the parallel direction. The two commonly used closures are sheath dissipation
closure, neglecting parallel gradients, or the vorticity advection closure,
neglecting parallel currents. Both cannot reproduce the results from full
3D simulation~\cite{easy14a}. Boltzmann spinning and the associated poloidal motion
is also not observed with 2D closures~\cite{easy14a}. Further drift waves
cannot be captured properly by 2D simulation~\cite{angus12a}.
Going towards a more complete picture of the physics, the complexity
of simulations was further increased.
For example finite thermal perturbation can significantly influence
filament dynamics, as it increases the poloidal motion and decreases
the radial velocity~\cite{walkden16a}.
For a more complete review of both the computational advances, as well
as experimental observations, see the review given
in~\cite{dippolito11a}.

Neutral plasma interactions are important for the operation of fusion
devices, in particular for detached operation, where neutrals
dissipate the majority of the parallel heat fluxes in the divertor
region. Also in the 
attached regime, neutrals can have a significant influence on plasma
dynamics.
Compared to present day machines, future fusion devices will have an
increased density in the divertor. This further increases the
importance of understanding the influence of neutral plasma
interactions in the divertor.
Plasma wall interaction are a main issue, not only for the
operation of ITER, but also for future fusion devices. Therefore
increasing the understanding of filaments, one of the main transport
mechanisms in the SOL is needed, especially in the
presence of neutrals.
Plasma turbulence interactions with neutrals studies have
recently been conducted~\cite{leddy16a,bisai16a,wersal15a}.
Leddy et al~\cite{leddy16a} show that the neutral interaction can be increased by
resolving the fluctuations, compared to the mean field
approach. Bisai and Kaw~\cite{bisai16a} show that neutrals can reduce
electric fields, reduce fluctuations and increase pressure gradients
in the SOL.
In terms of filament neutral interaction, a recent study
showed that filaments can significantly increase the fuelling of the
core by creating energetic neutrals~\cite{thrysoe16a}.
Scaling laws, describing the filaments radial velocity as a function of
plasma background parameter, have been
derived~\cite{garcia06a,walkden16a,yu06a}.
The scalings neglect not only neutral plasma interactions, but
simplify the equations in further ways, to get an analytical
expression for the filament velocity.

The study presented here extends this by taking not only the plasma
neutral interaction into account, but further looking at self
consistent parallel background profiles which include parallel
gradients. By looking at both direct and indirect interactions between
filaments and the neutral population,
these simulation extend the earlier study of direct
interactions~\cite{schworer17a}.
By looking at the influence of background profiles, earlier studies
looking at the influence of resistivity are extended in a self
consistent way~\cite{easy16a}.

The equations and the setup used here are described in
section~\ref{s:eq}, followed by a short introduction to the
background profiles in section~\ref{s:bg}. The 3D dynamics of the
simulated filaments is discussed in section~\ref{s:fil}. This is followed
by the influence of the neutrals, the background profiles in general,
and the filament size in section~\ref{s:res}, before the summary in
section~\ref{s:sum}.


\section{Modelling setup}\label{s:eq}
The model is based on the STORM module~\cite{easy14a,easy16a,walkden16a,schworer17a}, using
BOUT++~\cite{dudson09a,dudson15a}.
In this section we first discuss the simplified 3D straight field line
SOL geometry and then we present the drift ordered fluid equations.

The direction along the magnetic field is denoted by $z$.
The target is at $z=\pm L_z = \pm L_\parallel$, where sheath boundary
conditions are enforced.
Due to the symmetry of the system, only half of the domain is
simulated, namely $z = [0,L_z]$.
At $z=0$ symmetry boundary conditions are applied.

In addition to the parallel direction, the domain is spanned by
the radial direction
denoted by $x$ and the bi-normal direction, denoted by $y$.
The length along the magnetic field is $L_z=10$\,m, and is resolved by
$n_z=64$ grid points.
In the perpendicular direction the length are $L_x=L_y=10\delta_\perp$
i.e. dependent on the perpendicular extent of the filament
$\delta_\perp$.
The resolution is $n_x=n_y=128$. For $\delta_\perp=20$\,mm this gives
a grid spacing if $dx=1.5625$\,mm.
The filament size of $\delta_\perp = 20$\,mm was chosen, as it is
both close to the critical size $\delta^*$ (introduced later), but
also similar to the 
size experimentally observed in MAST\cite{schworer17a,walkden16a}.

The STORM model is a drift ordered full fluid model, following the
approach of Simakov and Catto~\cite{simakov03a,simakov04a}.
The equations are given in Bohm units~\cite{walkden16a}. The time is
normalized using the ion gyro frequency $\Omega_i$, lengths with the gyro
radius $\rho_s=c_s/\Omega_i$ and speeds with the speed of sound
$c_s=\sqrt{T_e/m_e}$.

The model consists of the electron density $n$ continuity equation
\begin{align}
     \frac{\partial n}{\partial t} &=
          \conv{\frac{\nabla \phi \times \vek b}{B} \cdot \nabla n}
          - \conv{\nabla_\parallel (V n)}
          + \diff{\mu_n \laplace n}
          - \curv{g n \frac{\partial \phi}{\partial y}}
          + \curv{g \frac{\partial n T}{\partial y}}
          + \neut{\Gamma^\text{ion}}
          - \neut{\Gamma^\text{rec}}
          \\[0.5ex]
          \intertext{with the potential $\phi$ being the Laplacian
          inversion $\omega=\laplace_\perp \phi$ of the vorticity.
          $B$ is the magnitude of the magnetic field, and $\vec b$ is
          its direction.
          $\mu_n$ is the diffusion coefficient for the electron
          density.
          The terms with $g$ are terms due to curvature, which are
          artificially reintroduced, to drive the filaments. $g$ is
          related to the radius of curvature $R_c$ as
          $g=\frac{2}{R_c}\approx 1.33$\,m$^{-1}$.
          $\Gamma^\text{ion}$, $\Gamma^\text{rec}$ and $\Gamma^\text{CX}$
          are the ionization, recombination and charge exchange rates.
          The equation for the parallel electron
          velocity $V$ is}
     \frac{\partial V}{\partial t} &=
          \conv{\frac{\nabla \phi \times \vek b}{B} \cdot \nabla V}
          - \conv{V \nabpara V}
          + \esta{\mu \nabpara \phi}
          - \pres{\frac{\mu}{n}\nabpara nT}
          \\
          &\quad
          + \fric{n \mu \eta_\parallel (U-V)}
          - \tpho{0.71 \mu\nabpara T}
          - \neut{\frac V n \Gamma^\text{ion}}\nonumber
          \\
          \intertext{with the ion-electron mass ratio is
          $\mu=m_i/m_e$. The parallel ion-electron resistivity is
          given by $\eta_\parallel$. The equation for the parallel ion
          velocity $U$}
     \frac{\partial U}{\partial t} &=
          \conv{\frac{\nabla \phi \times \vek b}{B} \cdot \nabla U}
          - \conv{U \nabpara U}
          - \esta{\nabpara \phi}
          - \fric{\eta_\parallel n (U-V)}
          \\
          &\quad
          + \tpho{0.71 \nabpara T}
          - \neut{\frac{U} n \Gamma^\text{ion}}
          - \neut{\frac U n \Gamma^\text{CX}}\nonumber
          \\
          \intertext{the equation for the electron temperature $T$}
     \frac{\partial T}{\partial t} &=
          \conv{\frac{\nabla \phi \times \vek b}{B} \cdot \nabla T}
          - \conv{V \nabpara T}
          + \frac{2}{3} \left( \frac{-1}{n}\nabpara{q_\parallel}
            + \tpho{0.71 (U-V) \nabpara T}
            - T \nabpara V\right. \label{eq:temp}
          \\
          &\quad\left.
            + \diff{\frac{\kappa_\perp}{n} \laplace_\perp T}
            + \fric{\eta_\parallel n (U-V)^2}
            \right )
          - \curv{\frac 2 3 g T \frac{\partial \phi}{\partial y}}
          - \curv{\frac 2 3 g \frac{T^2}{n} \frac{\partial n}{\partial y}}
          - \curv{\frac 7 3 g T \frac{\partial T}{\partial y}}\nonumber
          \\
          &\quad- \curv{\frac 2 3 g V^2 \frac{1}{\mu n} \frac{\partial n T}{\partial y}}
          - \neut{\frac T n \Gamma^\text{ion}}\nonumber
          \intertext{The parallel heat conduction is
          given by $q_\parallel$ and $\kappa_\perp$ is the
          perpendicular heat transport coefficient.
          The equation for the vorticity $\omega$ is}
     \frac{\partial \omega}{\partial t} &=
          \conv{\frac{\nabla \phi \times \vek b}{B} \cdot \nabla \omega}
          - \conv{U \nabla_\parallel \omega}
          + \esta{\nabla_\parallel (U-V)}
          + \esta{\frac{U-V}{n}\nabla_\parallel n}
          + \diff{\mu_\omega \laplace \omega}\label{eq:vort}
          \\
          &\quad
          + \diff{\nabla_\perp \mu_\omega \cdot \nabla_\perp \omega}
          + \curv{\frac g n \frac{\partial n T}{\partial y}}
          - \neut{\laplace_\perp \phi (\Gamma^\text{CX} + \Gamma^\text{ion})}
          - \neut{\nabla_\perp \phi \cdot \nabla(\Gamma^\text{CX} + \Gamma^\text{ion})}\nonumber
          \\
          \intertext{with the vorticity diffusion coefficient $\mu_\omega$.
          The equation for the neutral density $n_n$ is }
     \frac{\partial n_n}{\partial t} &=
          \diff{\nabla (D_n \nabla n_n)}
          - \neut{\Gamma^\text{ion}}
          + \neut{\Gamma^\text{rec}}
          + \neut{S_R}
          - \neut{f_l n_n}
\end{align}
The diffusion constants, resistivity and neutral rates are calculated self
consistently~\cite{walkden16a}.
%
%
In the equation for the neutral density,
$D_n$ is the neutral diffusion, given by
\begin{align}
  D_n^0 &= \frac{v_{th}^2}{v_{th}\sigma n_n + \Gamma^\text{CX} +
    \Gamma^\text{ion}}\\
  D_n &=\begin{cases}
          D_n^0 &\mbox{if }D_n^0 \geq 2D^0\\
          D_n^0/2+D^0 &\mbox{if }D_n^0 < 2D^0\\
  \end{cases}
\end{align}
with $v_{th}$ deuterium's thermal speed at 300\,K and the atomic
deuterium-deuterium cross section $\sigma=\pi (52.9$\,pm$)^2$.
The diffusion limiter $D^0$ is needed to compensate for the lack
of pressure in high neutral density regions, in which case an
unphysically high diffusion occurs.
The term $f_l n_n$ emulates cross field losses.
Recycling of the neutrals is proportional to the particle flux at the target
$f_T=nU|_\text{target}$, the
recycling coefficient $f_R=0.9$ and depends on a Gaussian
recycling falloff length $L_R=1$\,m:
\begin{align}
  S_R &= \alpha_R \frac{f_R} f_T \exp(-z^2/L_R^2)
\end{align}
where $\alpha_R$ is a normalization constant, ensuring that a fraction
$f_R$ of the target flux $f_T$ are recycled along the field line.
This non-local model was chosen, as the lack of pressure combined
with high neutral densities near the target results in low return fluxes
of particles back along the field line. 
This non-local recycling model combined with a limiter for the
neutral diffusion $D_n$ ensures that the neutral are transported from
the target up stream.
The recycling model is an extension of the density source
previously used in STORM~\cite{easy14a,easy16a,walkden16a}.

In the radial direction Neumann boundary conditions with zero gradient are
enforced, with the exception of $\omega$ and $\phi$, which are set to
respective background values.
The $y$ direction is periodic for all quantities.
At the symmetry plane the velocities $U$ and $V$ are set to zero,
whereas for the other quantities zero gradients are enforced.
At the target magnetic pre-sheath boundary conditions were set.
The ions need to reach the speed of sound  $U=\sqrt{T}$, and the
electrons have to reach at the sheath boundary
\begin{align}
  V &= \sqrt{T} \exp(-V_f-\frac \phi T)
\end{align}
where $V_f$ is the floating potential~\cite{chodura81a,stangeby95a}.
The neutral density is forced to have a vanishing gradient at the
target boundary.

\section{Background profiles}\label{s:bg}
In order to study the influence of self consistent backgrounds on
filaments, a procedure for producing such backgrounds is needed.
Filaments will be seeded on these backgrounds, as described in
section~\ref{s:fil}.
The backgrounds are an extension of the two-point model.
They feature only dependence along the magnetic field, and
not in the radial direction.
In order to generate the one dimensional background profiles, the
equations presented above were used, with the
perpendicular terms dropped and the current forced
to be zero. The particle and energy influx was set to an
exponential shape, to localize the influx at the mid-plane.
The magnitude was controlled with an PID-controller to
achieve a predefined value for upstream temperature and density.
A PID controller sets the influx as a function of the instantaneous
difference to the predefined value, the integral and the derivative of
the difference. It is a commonly used control loop feedback mechanism.
For
the 3D simulation the controller is replaced by the steady state value
of the background simulation.
While setting the value via a Dirichlet boundary condition would be
easier in the case of the background profiles, the influx needed for
maintaining the background isn't known. This causes issues for the
filament simulations, as a Dirichlet boundary condition 
would interact non-trivially with the seeded filament. Further, a Dirichlet
boundary condition would concentrate all the influx in a single point,
instead of spreading it.

\begin{figure}
  \centering
  \includegraphics[scale=1.1]{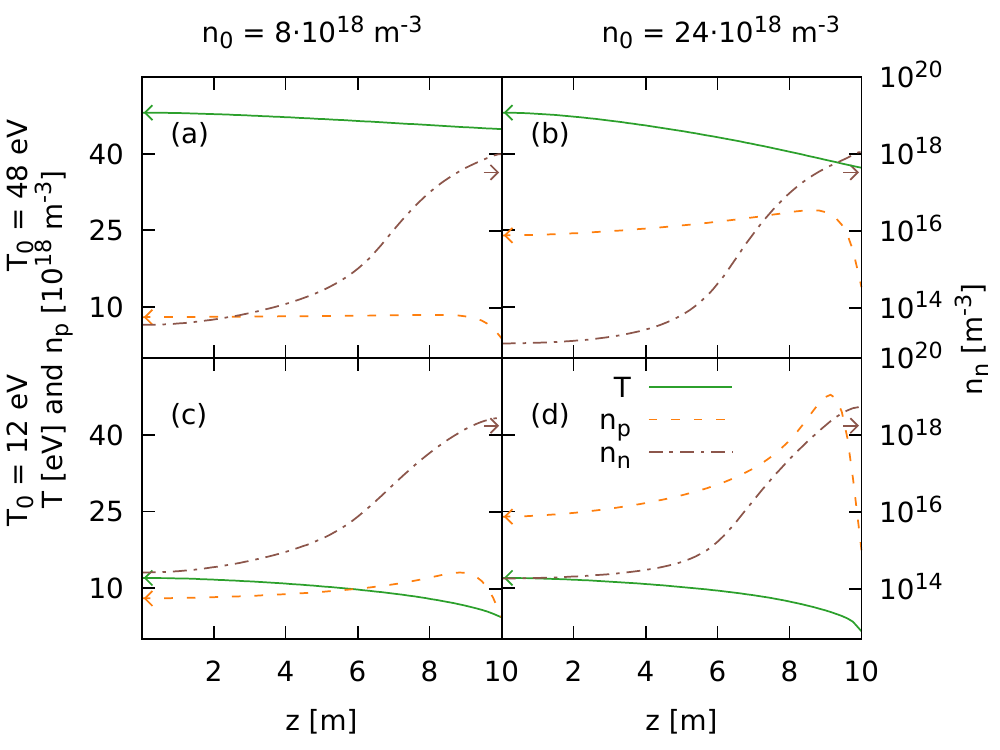}
  \caption{Background plasma profiles, run to steady-state for a set
    upstream temperature $T_0$ and density $n_0$. The sheath is at the right hand
    side at $L=10$\,m. The mid-plane is at the left side, and is a
    symmetry plane.
    The profiles (a) and
    (b) have an upstream electron temperature of $T_0=48$\,eV at the mid-plane,
    while (c) and (d) have a upstream electron temperature of
    $T_0=12$\,eV. The upstream background density for (a) and (c) is
    $n_0=8\times10^{18}$\,m$^{-3}$, for (b) and (d)
    $n_0 = 24 \times 10^{18}$\,m$^{-3}$ at the mid-plane. Plasma
    density and temperature is plotted to the linear scale on the
    left hand side.
    The neutral density $n_n$ is plotted to the log scale on the right
    hand side.}\label{f:bg}
\end{figure}
In order to generate different profiles, the upstream electron
temperature $T_0$ and upstream electron density $n_0$ were scanned,
which allowed for different SOL regimes
to be investigated. Fig.~\ref{f:bg}
shows temperature and density of the electrons, as well as neutral
density.
The 12\,eV temperature simulations are in the high recycling regime,
as the temperature drops significantly along the field line.
The high temperature simulations are in the low recycling regime (also known
as the sheath-limited regime)
~\cite{stangeby00a}. Note that these simulations do not feature
detachment, which requires a more precise treatment of neutrals.

\begin{figure}
  \centering
  \includegraphics[scale=1.1]{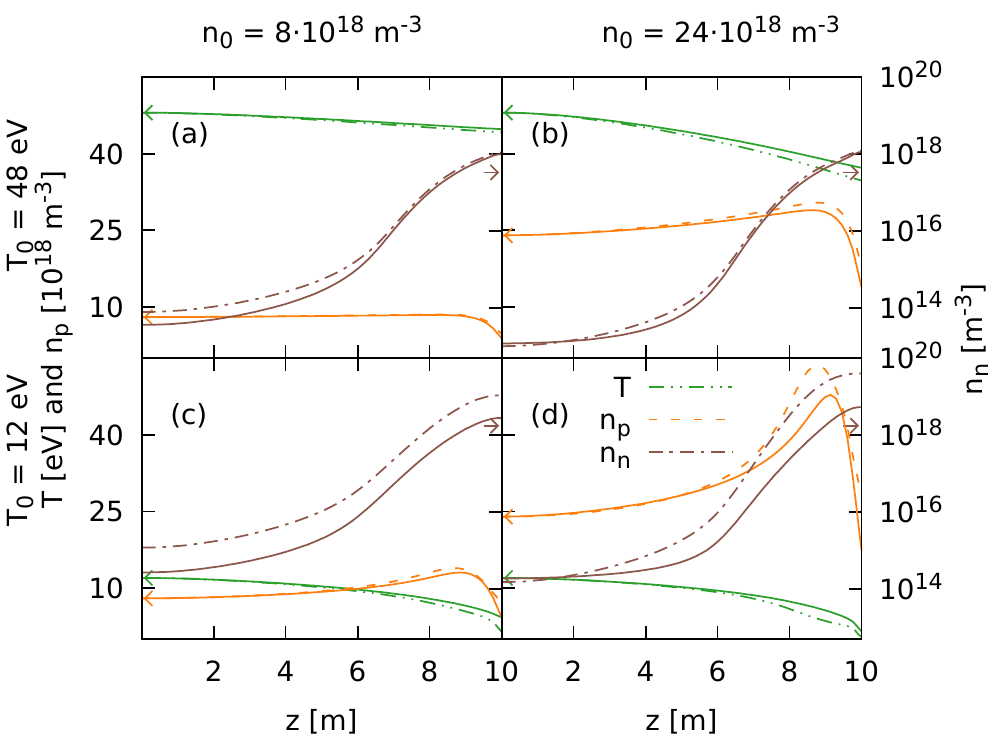}
  \caption{Background plasma profiles, see fig.~\ref{f:bg} for a
    detailed description.
    Additional to the profiles of the full neutral model, shown using
    continuous lines, the ionization-only backgrounds are
    shown using dashed lines.}\label{f:bgoo}
\end{figure}

In order to reduce the interaction of the filaments with neutrals,
a second set of background profiles was generated, where the plasma
neutrals interaction was limited to ionization. This allows us to keep
the recycling dominated fuelling of the plasma, without the need to change
the model, except setting $\Gamma^\text{CX}=\Gamma^\text{rec}=0$.
Fig.~\ref{f:bgoo} shows that in this case, the temperature decreases
faster towards the target. This leads to slower ion and electron
velocities at the target.
Further, the densities of both plasma and neutrals is increased.
The strongest differences between the models is in the low
temperature cases.

\section{Filament evolution}\label{s:fil}
The filament is seeded as a density and temperature perturbation on top
of the background profiles.
The initial shape in the perpendicular direction is Gaussian. The width
$\delta_\perp$ is, unless otherwise noted, 20\,mm.
In the parallel direction a $\tanh$ shape is used with a typical
parallel length of 5\,m.

Theoretical predictions suggest a scaling of the radial velocity of
the filament, that scales with
\begin{align}
  v_r^{s} &\propto \frac{\delta_p\sqrt{T_0+\delta_T}}{n_0+\delta_n}\label{eq:vr_s_f}
  \intertext{for the sheath limited regime and}
  v_r^{i} &\propto \sqrt{\frac{\delta_p}{n_0+\delta_n}}\label{eq:vr_i_f}
\end{align}
for the inertial limited regime \cite{walkden16a}.
$\delta_\alpha$ is the perturbation above the background value
$\alpha_0$, for $\alpha\in {n,T}$.
The pressure perturbation $\delta_p$ consists of density and
temperature perturbation
$\delta_p=\delta_n T_0+ \delta_Tn_0+\delta_T\delta_n$.
To simplify the scalings, we take a density perturbation
$\delta_n$ equal to the upstream density, $n_0$, such that
$\frac{\delta_n}{n_0}=1$. Doing the same for 
the temperature perturbation, setting $\delta_T=T_0$ yields for the
pressure perturbation $\delta_p=3n_0T_0$. The scalings~\eqref{eq:vr_s_f}
and \eqref{eq:vr_i_f} reduce to:
\begin{align}
  v_r^{s} &\overset{\delta_\alpha=\alpha_0}{\propto} \frac{n_0T_0\sqrt{T_0}}{n_0}\\
  v_r^{i} &\overset{\delta_\alpha=\alpha_0}{\propto} \sqrt{\frac{n_0T_0}{n_0}}
\end{align}
yielding a temperature dependence of $T_0^\frac{1}2$ for the inertial
regime and $T_0^\frac{3}2$ for the sheath limited regime and no
dependence on the density.
This convention for the filament perturbations will be adopted
through out this paper. Within the scaling $T_0$ describes a
``background'' temperature. As the background temperature changes
along the magnetic field lines, it is not obvious how this $T_0$ for
the scaling should be calculated.

From the filament simulations the centre of mass was
calculated in the radial direction $c_r$:
\begin{align}
  c_r &=\frac{ \int\int x \Delta n(x,y,z) \ud x \ud z}
  {\int\int \Delta n(x,y,z) \ud x \ud z}
  \intertext{with}
  \Delta n(x,y,z)&= \begin{cases} n(x,y,z) - n_\text{cut}(z)&
    \text{for~} n(x,y,z) - n_\text{cut}(z) > 0 \\
    0 & \text{else}
    \end{cases}
\end{align}
where the cut of density $n_\text{cut}$ was computed by taking the
background density of that cross-section.
As the initial amplitude near the target is very small, the shown
results are measured near the mid-plane.
However the filaments move rigidly, so this velocity is representative
of the whole filament.

For each filament simulation, the maximum of the centre-of-mass
velocity is computed and compared.

\begin{figure}
  \centering
  \includegraphics[width=.9\linewidth]{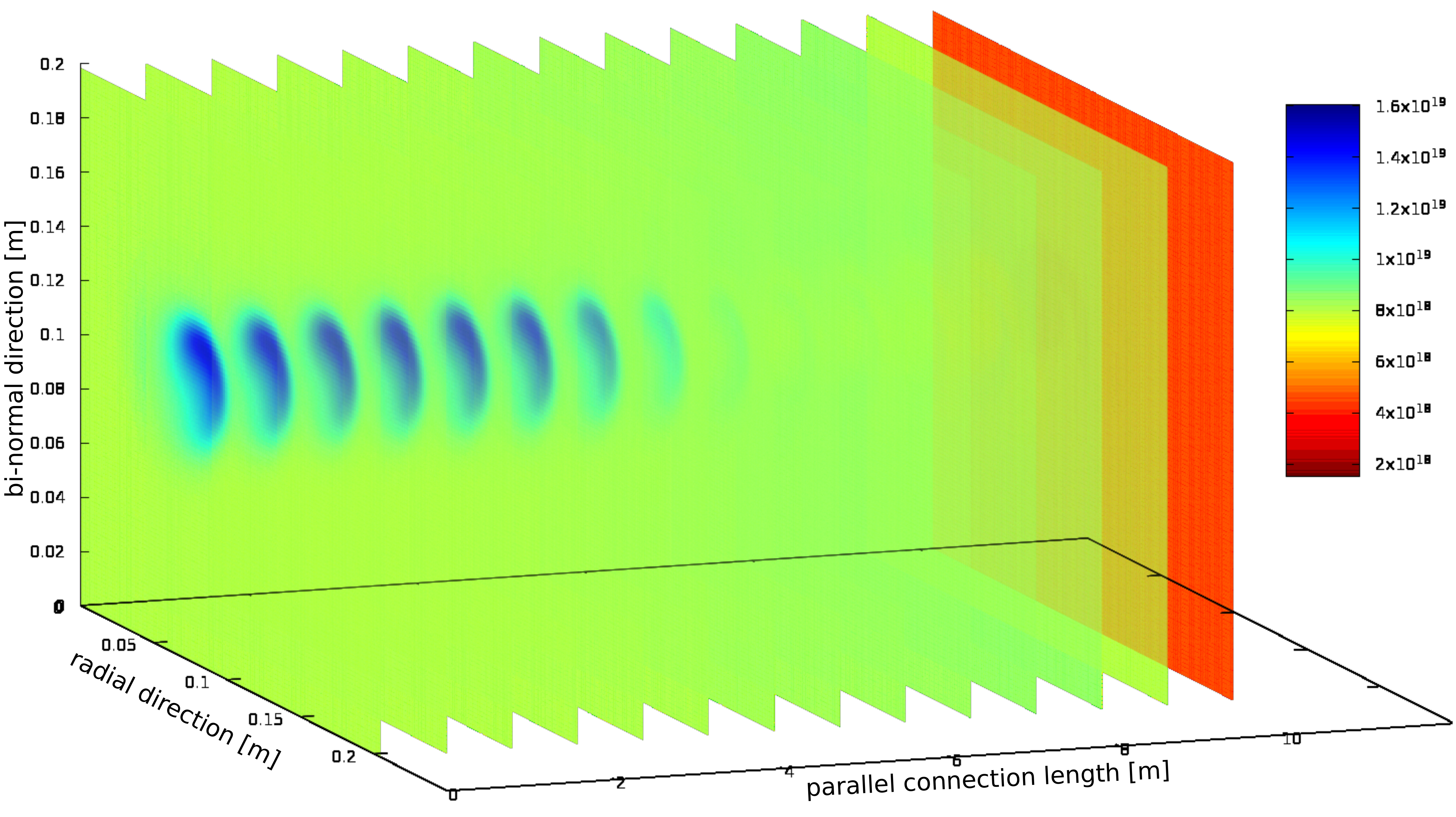}
  \caption{Snapshot of the density
    $\sim 7.5$\,µs after the filament was seeded.
    The upstream background plasma density was
    $n_0=8\times 10^{18}$\,m$^{-3}$ and the upstream electron
    temperature was $T_0 = 48$\,eV. The perpendicular
    size of the filament was $\delta_\perp=20$\,mm.}\label{f:evo}
\end{figure}
An example of a filament shape is shown in fig.~\ref{f:evo}.
The mushrooming behaviour, typical for these
filaments~\cite{dippolito11a,easy14a,walkden16a}, can be seen.
The filament is not symmetric in the $y$-direction. This motion in the
$y$-direction, due to the temperature perturbation, has
been observed and discussed~\cite{walkden16a,myra04a}.
The temperature perturbation causes an even parity contribution in the
potential due to the sheath potential, which causes Boltzmann
spinning.

\section{Influence on filament velocity}\label{s:res}
In order to distinguish the direct and the indirect influence of
neutrals on filaments, first different neutral models are compared.
This will be followed by a study of the background dependence on
filaments, before we conclude with results of the filament size
dependence.

\subsection{Neutrals}
In order to study the direct interaction between the neutrals and the
filaments, different neutrals-filament interaction models were used.
\begin{figure}
  \centering
  \includegraphics[scale=0.98]{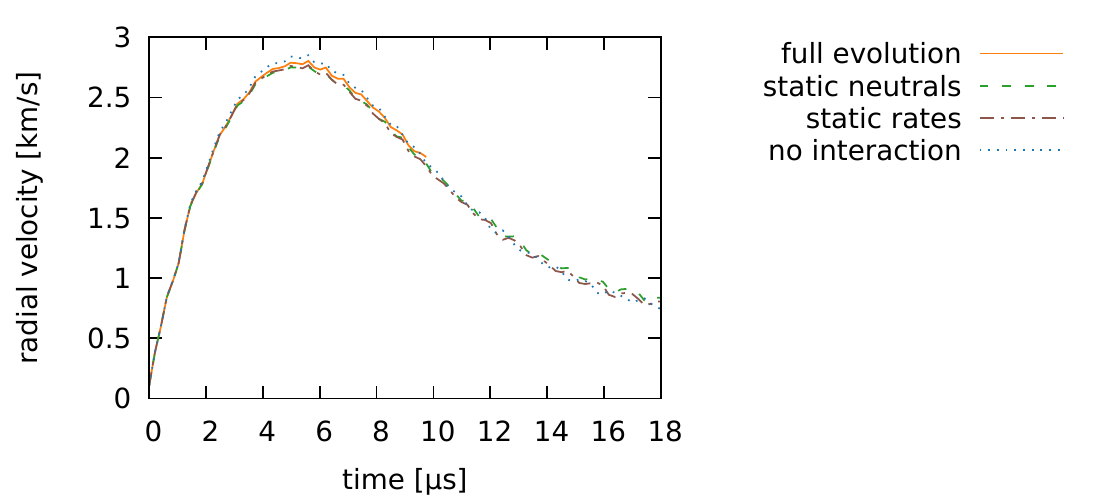}
  \caption{Comparison of different neutral filament interaction
    models. In the full evolution case, the neutral density was
    co-evolved with the filament, and the rates were calculated self
    consistently. In the no interaction case, the neutral term in the
    vorticity equation was switched of.
  In the static-rates case the neutral interaction rates
  $\Gamma^\alpha$ from the equilibrium profiles were used.
  In the static neutrals
    case, the neutrals were not evolved, but the rates were
    calculated.}\label{f:direct}
\end{figure}
The results are shown in fig.~\ref{f:direct}.
The no interaction case is where the neutral term in the vorticity
equation is set to zero.
For the other terms the neutrals are kept static and the neutral
rates are calculated self consistently.
In the static rates
case the neutral plasma interaction rates, namely charge exchange rate
$\Gamma^\text{CX}$, ionization rate $\Gamma^\text{ion}$ and
recombination rate $\Gamma^\text{rec}$, are kept at their steady
state values. This ensures that areas not affected by the filament are
kept at the steady state value.
The static rates case represents a state where the neutrals are still
interacting with the plasma, but the plasma filament interaction is
reduced to the neutral plasma background interaction.
The interaction is partially further switched on in the case of the static
neutrals. There the neutral profiles are not evolved, but the neutral
plasma interaction rates are calculated taking the filament into
account.
In the full evolution case the interaction is fully enabled. The
neutrals are evolved self consistently, and the interaction rates are
computed including both background and filament contributions to
density and temperature.

These simulations were done for the different backgrounds
shown in fig.~\ref{f:bg}.
The result shown in fig.~\ref{f:direct} is the one with the strongest
difference between
the velocities, the background profile with high density $n_0$ and high
temperature $T_0$.
It can be seen that there is only a small difference for the static
neutrals and static rates cases.
In the case where the neutrals are evolved, the filament moves slightly faster.
The filament is fastest when the neutrals drag term in the vorticity
equation is switched off. In the case where the neutrals are co-evolved
with the filament, the filament `burn` partially through the neutrals
which explains why the velocities lie between the static cases and
the no interaction case.

This shows that neutrals impact the motion of filament - at least in
the high density and high temperature cases.
The following results are obtained using the static neutral
approximation, as this significantly accelerates the computation.
The deviation from the full neutrals evolution is less then
1.5\,\%
in the conditions featured here,
which do not include detachment.

\begin{figure}
  \centering
  \includegraphics[width=.98\linewidth,page=2]{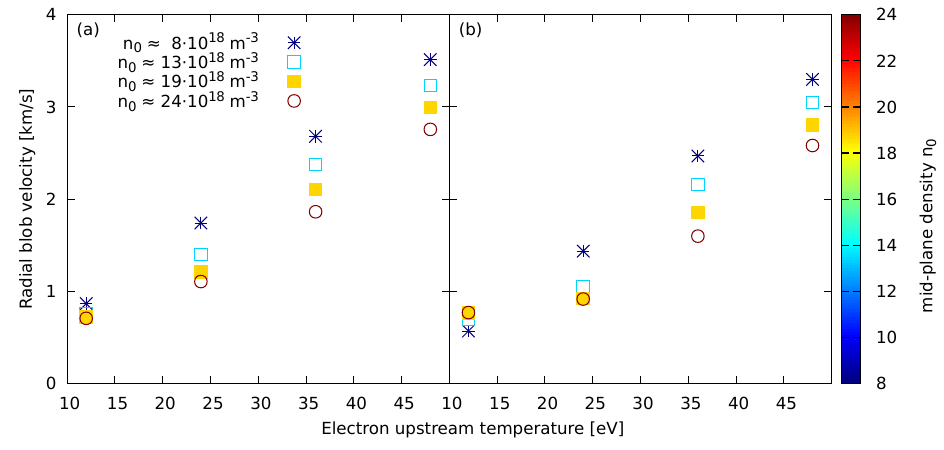}
  \caption{Comparison of the filament velocities for the different set
  of background profiles, (a) backgrounds with full neutral
  interaction and (b) without recombination and charge
  exchange.}\label{f:indirect}
\end{figure}
In addition to the weak dependence on the direct interaction between
filament and neutrals,
the filament velocity does vary with background conditions. This is
shown in fig.~\ref{f:indirect}, where on the left filaments were
seeded on the backgrounds with full neutral interactions.
Also shown is the effect of removing charge exchange and recombination
from the simulations. This impacts filament velocity through the
change in the backgrounds, indicating that neutrals are important and
interact with the filament indirectly via the plasma background.
In the next section, the dependence of the filament velocities on the
background conditions is studied in more detail.

\subsection{Background dependence}
\begin{figure}
  \centering
  \includegraphics[width=.98\linewidth]{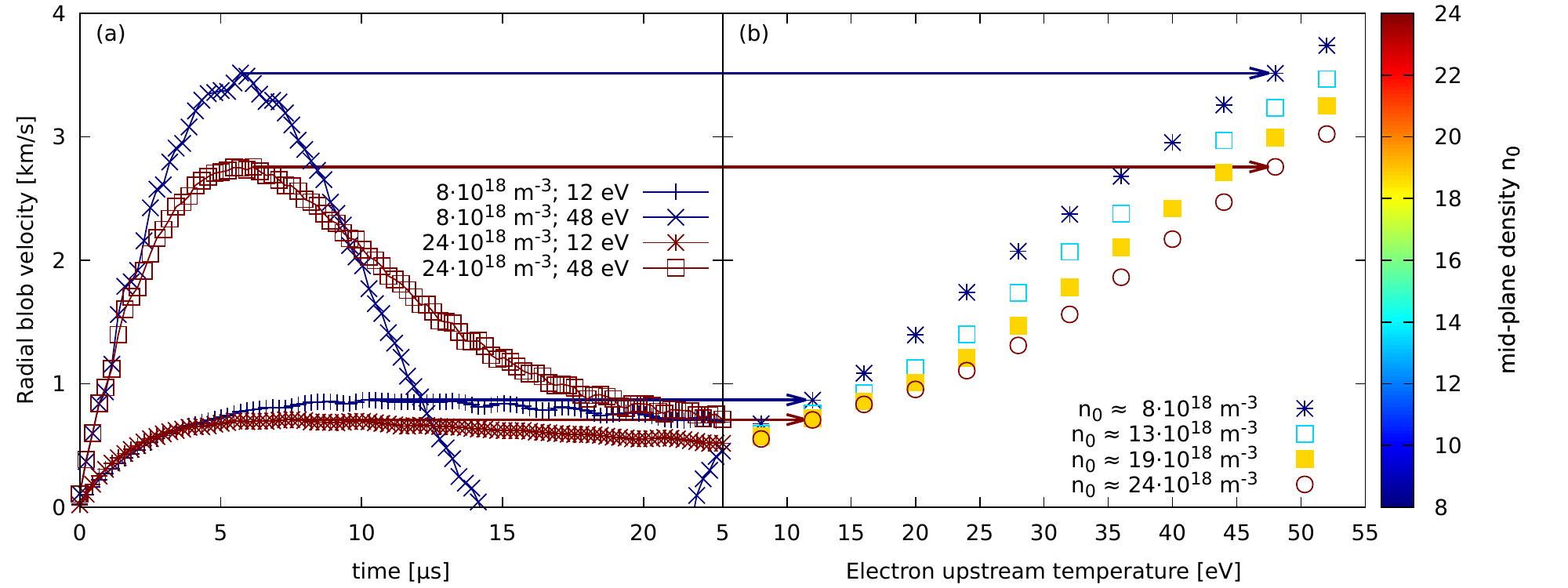}
  \caption{Radial velocity of filaments seeded on different
    backgrounds. Shown in (a) is the time evolution of the filaments,
    and (b) shows the peak radial velocities as a function of the
    upstream temperature.
  }\label{f:temp_up}
\end{figure}

This section presents the dependence of the filaments radial velocity
on the background conditions. Fig.~\ref{f:temp_up} (a) shows
the time evolution of filaments seeded on the
background profiles shown in fig.~\ref{f:bg}.
On the right hand side of fig.~\ref{f:temp_up} is a plot of the peak of the
filaments radial velocity as a function of the upstream temperature.
The velocity increases with an increase of temperature. The velocity
decreases with increasing density, with the exception of low
temperatures, where this trend is inverted. As the filaments are seeded such
that $\frac{\delta_n}{n}$ stays constant, the density dependence is not
expected from the simple scaling analysis shown in section~\ref{s:fil}.

Earlier studies in STORM looked at the influence of the
resistivity~\cite{easy16a}. This was done by artificially changing the resistivity. In
this study this is repeated in a self consistent way. In order to
change the resistivity, the temperature needs to be changed.
\begin{figure}
  \centering
  \includegraphics[width=.98\linewidth,page=2]{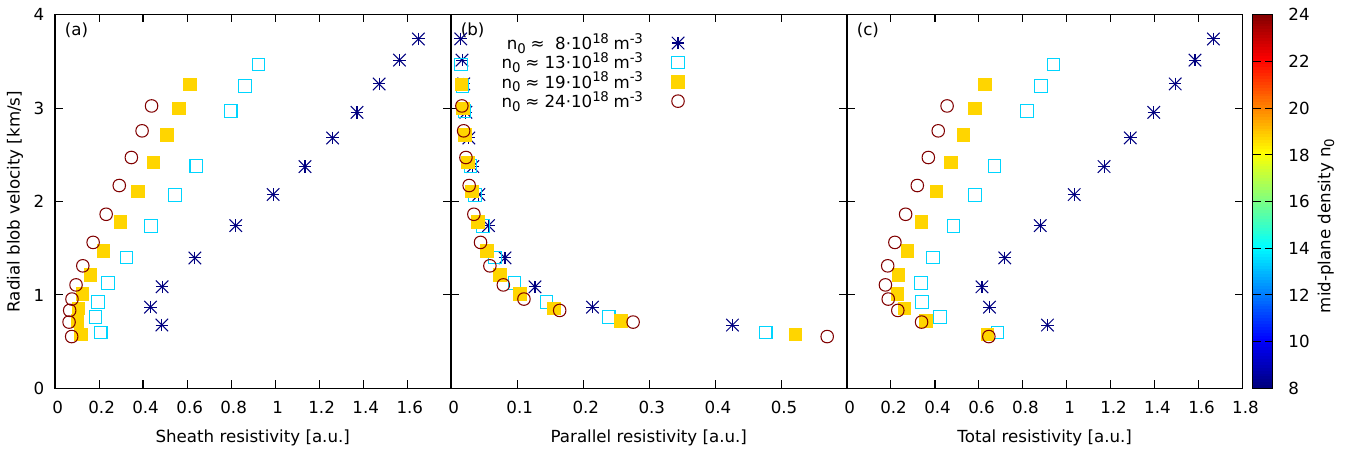}
  \caption{Radial velocity of filaments seeded on different
    backgrounds. Shown in (a) is the peak velocity as a function of
    sheath resistivity and (b) as a function of the parallel resistivity and
    (c) as a function of total resistivity.
  }\label{f:resi}
\end{figure}
Fig.~\ref{f:resi} shows the peak velocity as a function of (a) the
sheath resistivity, (b) the plasma resistivity integrated along the
magnetic field lines
and in (c) the total resistivity, consisting of the sum of both.
Note that the non-monotonic
behaviour below 1\,km/s is because the density at the target reduces
quite strongly with decreasing temperature, therefore the sheath
resistivity increases, and the colder temperatures have a higher
target resistivity. Fig.~\ref{f:resi} shows that the resistivity doesn't have a
major impact on the filament dynamics, and the temperature of the
filament is more important, for the conditions studied here.
As the plasma resistivity is a function of the temperature, the
scaling in (b) shows a monotonic decreasing behaviour. This is not
an effect of the resistivity, as with increasing resistivity, the
vorticity should increase, which would results in faster filaments \cite{easy16a}.
This shows that in this self consistent study, the change in
resistivity is less important than the associated change in
temperature.

\begin{figure}
  \centering
  \includegraphics[width=.99\linewidth,page=1]{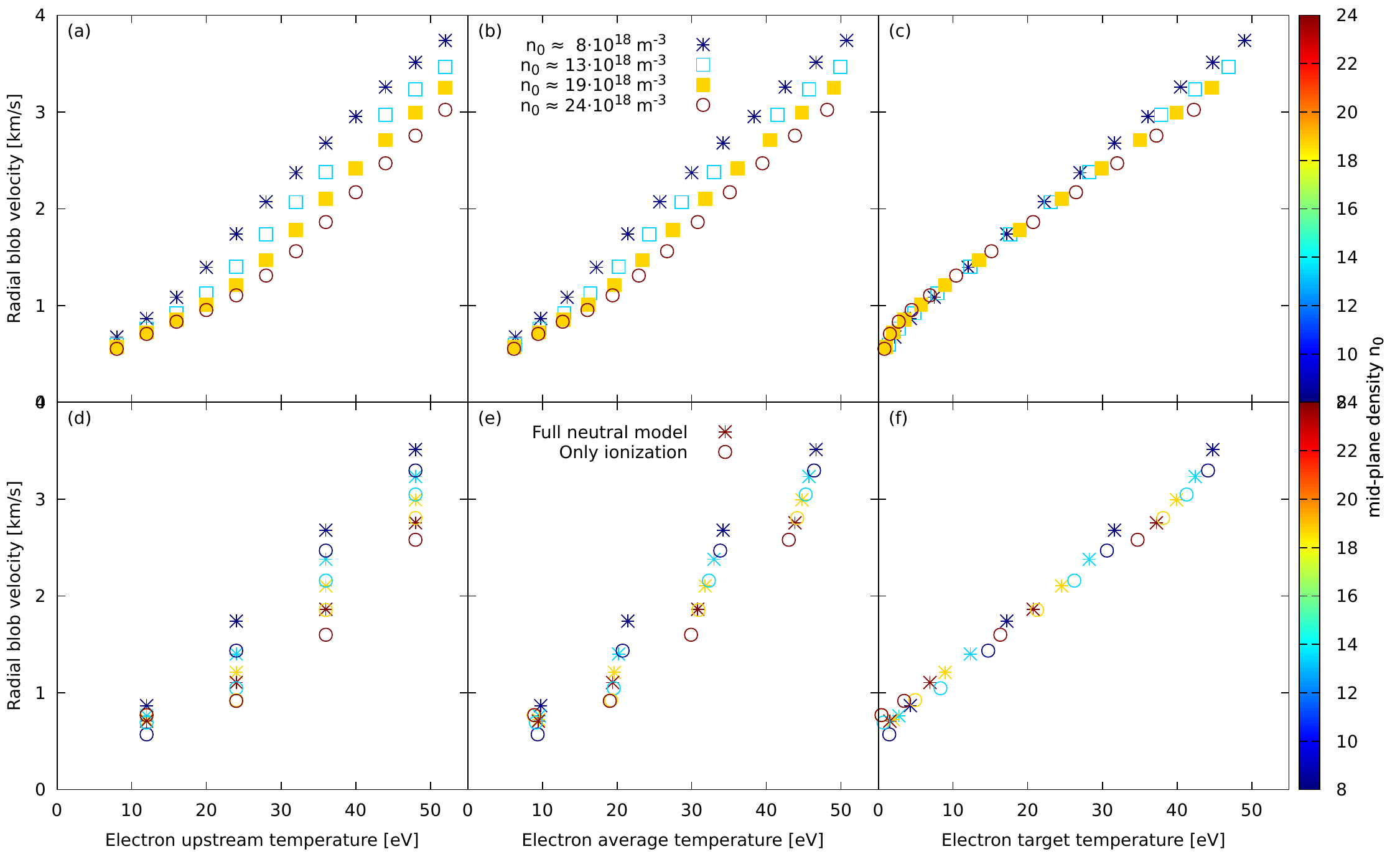}
  \caption{Radial velocity of filaments seeded on different
    backgrounds. Shown is the peak velocity as a function of
    the temperature. On the top (a-c) with static neutrals, while
    on the bottom (d-f) are the results for both the full neutral
    backgrounds  and the only ionization model.
    On the left (a,d) the velocity is plotted against the upstream
    temperature, in the middle (b,e) the velocity is plotted against
    the average temperature and on the left (c,f) the data is
    plotted against the target temperature.
  }\label{f:temp_t}
\end{figure}
The simple scaling analysis shown in section~\ref{s:fil} does require a
single background temperature, however the temperature is not constant
along the magnetic field lines.
Fig.~\ref{f:temp_t} shows the filament peak velocities for the
different profiles as a function of the upstream temperature, the
average temperature and the target temperature. In all cases a
monotonic increase with temperature is observed.
In the case of the target temperature,
the different upstream density profiles collapse approximately onto a single
line. This suggests that the target temperature is a good scaling
quantity for the radial velocity of the filaments studied here.
In fig.~\ref{f:temp_t} (f) the results from the different neutral
models are much closer to each other than in fig~\ref{f:temp_t} (d-e)
where they are plotted as function of the upstream temperature $T_0$
and the average temperature.

The vorticity equation, which determines the filament radial velocity,
represents a balance between parallel, polarization and viscous
currents with the driving diamagnetic currents in the filament.
As part of the filaments vorticity is closed via polarization
currents, we do not expect such a strong dependence on the target
temperature.
To study this further, a set of simulations was run,
removing the density dependence of the plasma viscosity $\mu_\omega$ in eq.~\eqref{eq:vort}.
$\mu_\omega$ has otherwise linear density dependence, so an increased
density leads to an increased diffusion of the vorticity, thereby
reducing the drive.
\begin{figure}
  \centering
  \includegraphics[width=.99\linewidth,page=2]{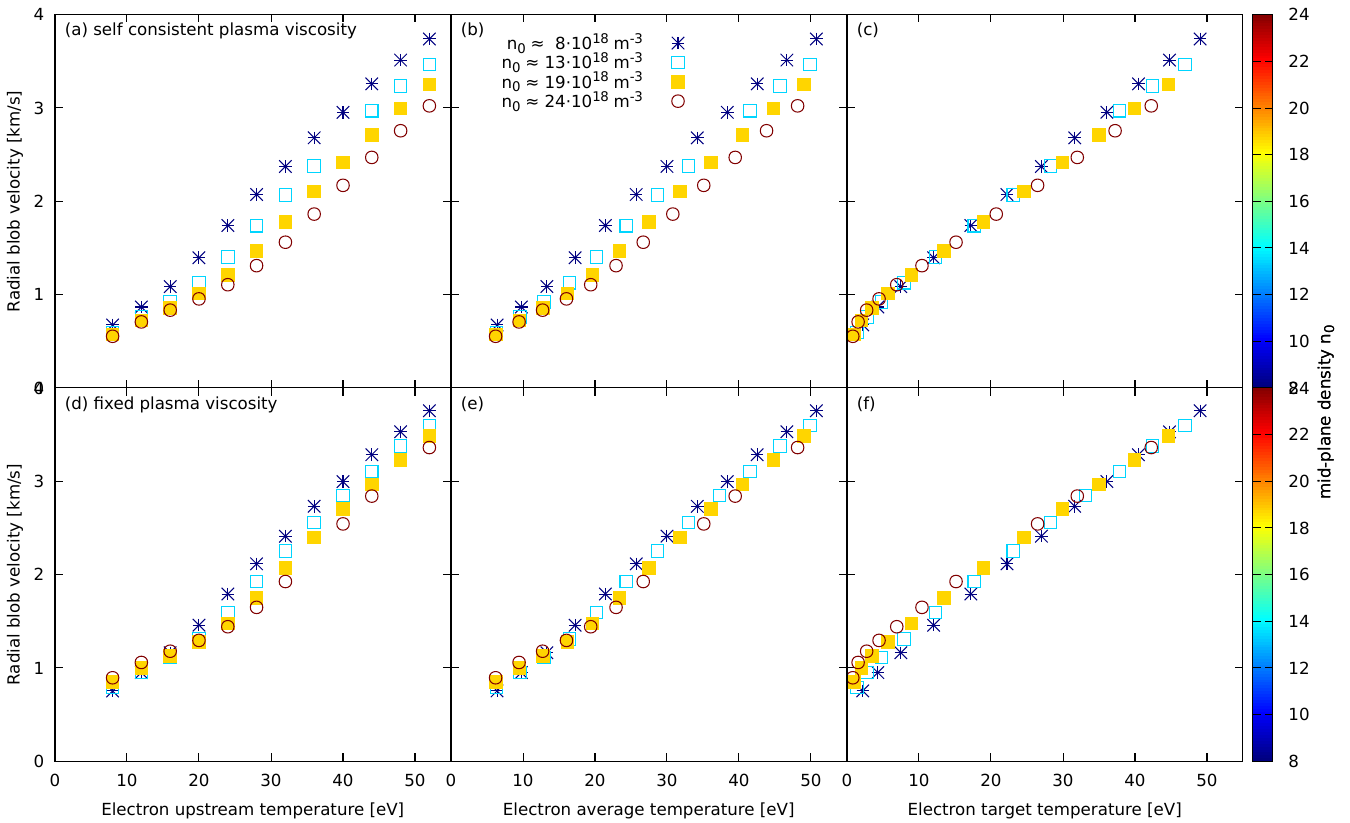}
  \caption{Peak radial velocity of filaments as a function of
    upstream, average and target
    temperature. The filaments in (a-c) are evolved with self
    consistent plasma viscosity and in (d-f) the density
  dependence of the plasma viscosity was removed.
  }\label{f:visc}
\end{figure}
Fig.~\ref{f:visc} compares the self consistent viscosity simulation
(a-c)
with the ones where the plasma viscosity has no density dependence (d-f). The density dependence, if
plotted against the upstream temperature, is
significantly reduced. In fig.~\ref{f:visc} (e) the radial velocity
is plotted against
the target temperature. Although the points do not collapse onto a
single 
line, they are still reasonably close to a single line.
This supports the point that the filaments aren't only influenced by
the target temperature.
It is worth noting that a similar data collapse is apparent in
fig.~\ref{f:visc}(d).
As there is no reason that filaments should be influenced by the
average temperature if the density dependence of the viscosity is
fixed, while being influenced 
by the target temperature in case of the full dynamics, this supports
the point that the collapse onto a single line is a coincidence, and
most likely will not be true for other conditions.  As in the here
presented simulations sheath currents play a significant role, the
results are not directly applicable to situations where they are
suppresed, for example in detached regimes.

Looking at fig.~\ref{f:visc} (a) and (d), removing the density
dependence of the plasma
viscosity reduces the dependence of the filament velocity on
the density.
For the remaining density
dependence, different reasons come into play.
The plasma viscosity still has a temperature dependence and for higher
densities the target temperature drops to lower values than for higher
densities, causing a higher viscosity near the target.
This shows that the filaments are indeed influenced by the conditions
at the target.
Note that the filaments have been seeded unconnected from the sheath,
but due to the fast electon motion, they still connect to the target,
and are therefore influenced by the plasma conditions at the target.

Another reason for the density dependence is via the neutrals.
As shown earlier for the higher density cases the neutrals cause a
larger reduction in filament velocity.
Further, the background parallel velocity is decreasing
with increasing density. The change in the parallel velocity is
stronger for lower temperatures. This can be explained by an
increasing importance of recycling in comparison to upstream density
fuelling.
Although this contribution is only small, it 
might explain to some extent the crossover at low temperatures, where
low densities
are slower then high densities. This cross over is observed in the
$\mu_\omega$ case (fig.~\ref{f:visc} (b)) and the ionisation 
only case (fig.~\ref{f:indirect} (b)), while in the full case the
density dependence is reduced.

Finally parallel currents are playing a significant role in the
generation of vorticity. The parallel
currents are affected by the sheath conditions, as they are flowing
through the sheath. Therefore also currents closer to upstream, are
influenced by the sheath temperature.


To further test the dependence of the filament velocity on the various
temperatures within the system, the upstream and target temperatures
have been partially decoupled from one-another. This has been achieved
by inserting an artificial heat sink localised near the target to
control the target electron temperature independent of the upstream
temperature.
\begin{figure}
  \centering
  \includegraphics[width=.98\linewidth,page=2]{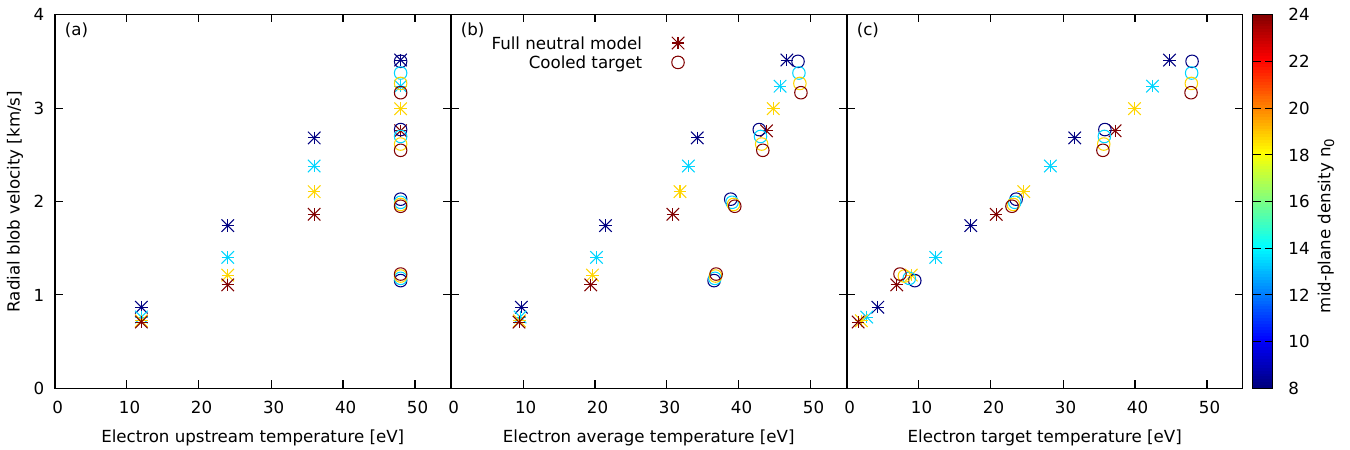}
  \caption{Peak radial velocity of filaments as a function of
    upstream and target
    temperature. In some of the simulations the plasma was cooled with
  and additional heat sink near the target.
  }\label{f:cool}
\end{figure}
This was done for the simulations with an upstream temperature of
$T_0=48$\,eV.
The temperature close the target was set to values between 12\,eV and
48\,eV. 
The radial velocities are shown in fig.~\ref{f:cool}.
Although the filaments were all seeded with the same perturbation of
$\delta_T=48$\,eV, the filament velocity agrees with the scaling
of the target temperature, rather then the upstream temperature.
Note that in this case the $\nabla n(U-V)$ term is significantly stronger
near the target, then in the simulations without the target heat sink.
Therefore the vorticity is larger in amplitude near the target, than
further upstream. Therefore the 
viscosity near the target has a strong influence, which results in a
strong influence of the target temperature.
As the target temperature also influences sheath currents, the strong
target temperature dependence of the filament velocity is probably due
to both the viscosity as well as the sheath currents.


\subsection{Filament size}
To study the influence of the size of the filament on its dynamics,
different sized filaments have been seeded, and their motion analysed.
\begin{figure}
  \centering
  \includegraphics[width=.98\linewidth,page=1]{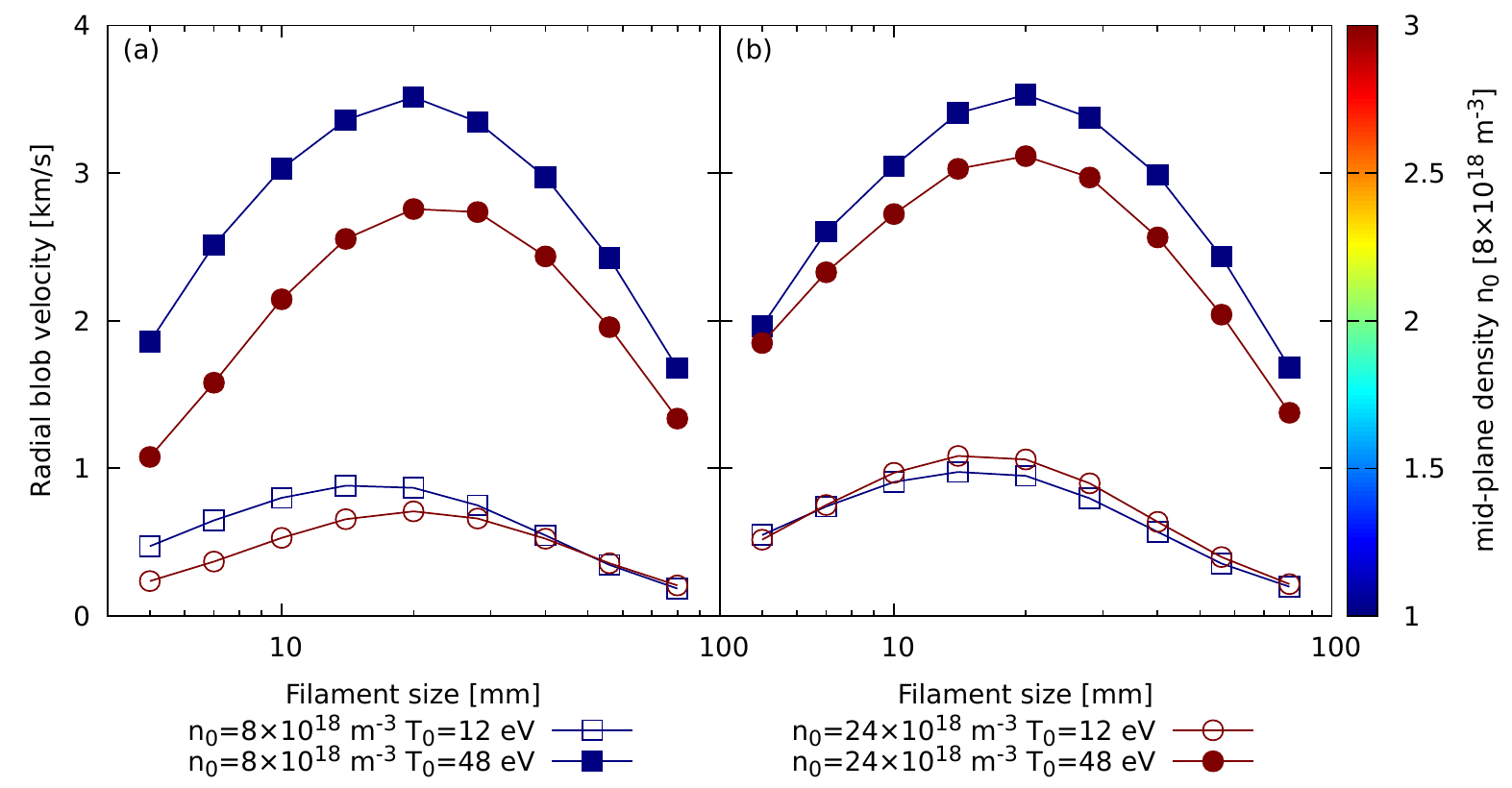}
  \caption{Radial velocity of different sized filaments.
    Shown is the peak velocity for the four backgrounds from
    fig.~\ref{f:bg}. On the left are simulation with the self
    consistent plasma viscosity, on the right hand side for 
    plasma viscosity without density dependence.
  }\label{f:size}
\end{figure}
Fig.~\ref{f:size} (a) presents the scan in filament size. It can be seen
that the filament size $\delta^*$, where the filaments are fastest
is for the $n_0=8\times 10^{18}$\,m$^{-3}$ between 14 and 20\,mm, and
for the $n_0=24\times 10^{18}$\,m$^{-3}$ between 20 and 28\,mm.
The position does seem to be only influenced by the upstream density
$n_0$, and not by the temperature.

As already done in the previous section, a scan where the plasma
viscosity had no density dependence was performed. This is shown in
fig.~\ref{f:size} (b).
In this case the fastest filaments are around
$\delta_\perp\approx 20$\,mm for the 48\,eV cases, and between 14 and
20\,mm for the 12\,eV case. This shows that the density dependence of
this point is due to the density dependence of the plasma viscosity,
which hasn't been included in past studies.
Further, a weak temperature dependence of $\delta^*$ is observed.
From the simple scaling derived in section~\ref{s:fil}, a temperature but
no density dependence is expected, suggesting that future derivations
of $\delta^*$ should include a self consistent plasma viscosity, and
currents due to viscosity.

The stronger density dependence of small filaments can be
explained by the density dependence of the viscosity.
As for small filaments the currents are closed via
currents in the drift plane, where viscous currents can contribute.
For large filaments, no dependence on the
viscosity is observed, as the currents are closed via the sheath.


\begin{figure}
  \centering
  \includegraphics[width=.98\linewidth,page=3]{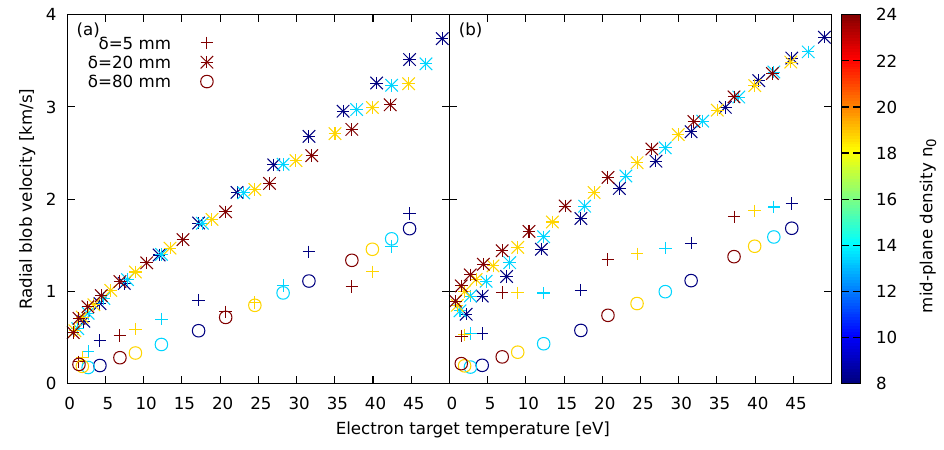}
  \caption{Radial velocity of different sized filaments.
    Shown is the peak velocity as a function of target temperature for
    different sized filaments. The $\delta_\perp=20\,mm\approx
    \delta^*$ are in size similar to the ones observed in MAST.
  }\label{f:temp_size}
\end{figure}
Fig.~\ref{f:temp_size} shows the peak radial velocity for different
sized filaments. The ones with size $\delta_\perp \approx \delta^*$
are the fastest ones.
The smaller ones and larger ones are significantly slower.
The larger ones collapse on a line. This agrees with theory, as the
vorticity for larger filaments is mainly closed via sheath currents,
therefore a dependence on the sheath conditions is expected.
The smaller ones, where the currents are closed mainly via
currents in the drift plane, show a stronger dependence on the density.
This strong density dependence can be explained by the viscosity.
If the density dependence of the viscosity is fixed, they do not collapse
that closely onto a single line, suggesting a weaker target dependence
compared to larger filaments.
As this geometry does not include an X-point, 
filaments can be connected to the target, and therefore influenced by
the target. If a more realistic geometry is used, it is quite likely
that at least for the smaller filaments the influenced of the
plasma condition at the sheath is reduced.

\section{Summary}\label{s:sum}
Filament radial velocities in the scrape-off layer for different
background profiles have been studied. Thereby the upstream
temperature and density have been varied, resulting in self consistent
parallel profiles. The backgrounds do not include gradients in the radial
direction.
Filaments were seeded on the background profiles, and the radial filament
velocity was measured.

It has been shown, that the direct interaction between the filament
and the neutrals are most strong in the high density and high
temperature case, where a weak reduction of velocity was observed.
The indirect interaction, via changing the background profiles have
been observed in all cases. 
To accurately capture filament dynamics, the parallel variation of the
background plasma, including interactions with the neutral population,
should be included.

Increasing the upstream temperature resulted in faster radial motion
of the filament, and decreased with increasing upstream density. This
can be explained by the reduced target temperature with increasing
density, as the target temperature was shown to be the best ordering
parameter for the filaments studied here.
As the filament perturbation is seeded unconnected to the sheath, it
is the fast electron motion, that connects the filament electricly to
the sheath.
This way the target temperature dependence can be explained by the
temperature dependence of the plasma viscosity and by sheath currents.

The strong target temperature dependence is not only observed for
filament sizes close to the 
critical size $\delta^*$ but also for larger ones. Here a
significant amount of the current is closed via sheath currents.
Smaller filaments show a strong dependence on plasma density, due
to the density dependence of the plasma viscosity.
If this influence is reduced, they show also a strong dependence on
the sheath temperature.
Further a shift of $\delta^*$ with density is observed. This is not
expected from scaling laws, but can be explained by the density
dependence of the plasma viscosity.
This suggests that the plasma viscosity
should be included if scalings for $\delta^*$ are derived.


The geometry used does not include an X-point or magnetic shear.
Furthermore detachment was not studied here, as a more accurate
neutral model would be required.
Both these aspects could reduce the target dependence, and further
studies are required to validate this findings in the case of detached
conditions or in scenarios including high magnetic shear.


\section{Acknowledgement}
This work has been carried out within the framework of the EUROfusion
Consortium and has received funding from the Euratom research and
training programme 2014-2018 under grant agreement No 633053 and from
the RCUK Energy Programme [grant number EP/P012450/1]. To obtain further
information on the data and models underlying this paper please
contact PublicationsManager@ukaea.uk. The views and opinions
expressed herein do not necessarily reflect those of the European
Commission.
Simulations in this paper made use of the ARCHER UK National
Supercomputing service (www.archer.ac.uk) under the Plasma HEC
Consortium EPSRC grant number EP/L000237/1.

\bibliographystyle{ieeetr}
\bibliography{phd}

\end{document}